\documentclass[showpacs,amsmath,amssymb,twocolumn]{revtex4}
\usepackage{graphicx}
\usepackage{bm}
\begin{document}
\title{Phenomenological model of protected behavior in the pseudogap state
of underdoped cuprate superconductors.}
\author{Victor Barzykin} 
\affiliation{Department of Physics and Astronomy, University of Tennessee,
Knoxville, TN  37996}
\author{David Pines}
\affiliation{MST-DO, MS G 754, Los Alamos National Laboratory, Los Alamos, NM 87545}
\affiliation{Department of Physics, University of California, Davis, CA 95616}
\begin{abstract}
By extending previous work on the scaling of low frequency magnetic properties 
of the 2-1-4 cuprates to the 1-2-3 materials, we arrive at a consistent phenomenological description 
of protected behavior in the pseudogap state of the magnetically underdoped 
cuprates. Between zero hole doping and a doping level of $\sim 0.22$  it reflects the 
presence of  a mixture of an insulating spin liquid that produces the measured 
magnetic scaling behavior and a Fermi liquid that becomes superconducting for doping levels $x>0.06$.  
Our analysis suggests the existence of two quantum critical 
points, at doping levels, $x \sim 0.05$ and $x \sim 0.22$,  and that d-wave superconductivity 
in the pseudogap region arises from quasiparticle-spin liquid interaction, i.e.  
magnetic interactions between quasiparticles in the Fermi liquid induced by 
their coupling to the spin liquid excitations.
\end{abstract}
\vspace{0.15cm}

\pacs{74.20.Mn, 75.40.Cx, 75.40.Gb, 76.60.-k}
\maketitle

\begin{figure}
\begin{center}
\includegraphics[width=3in]{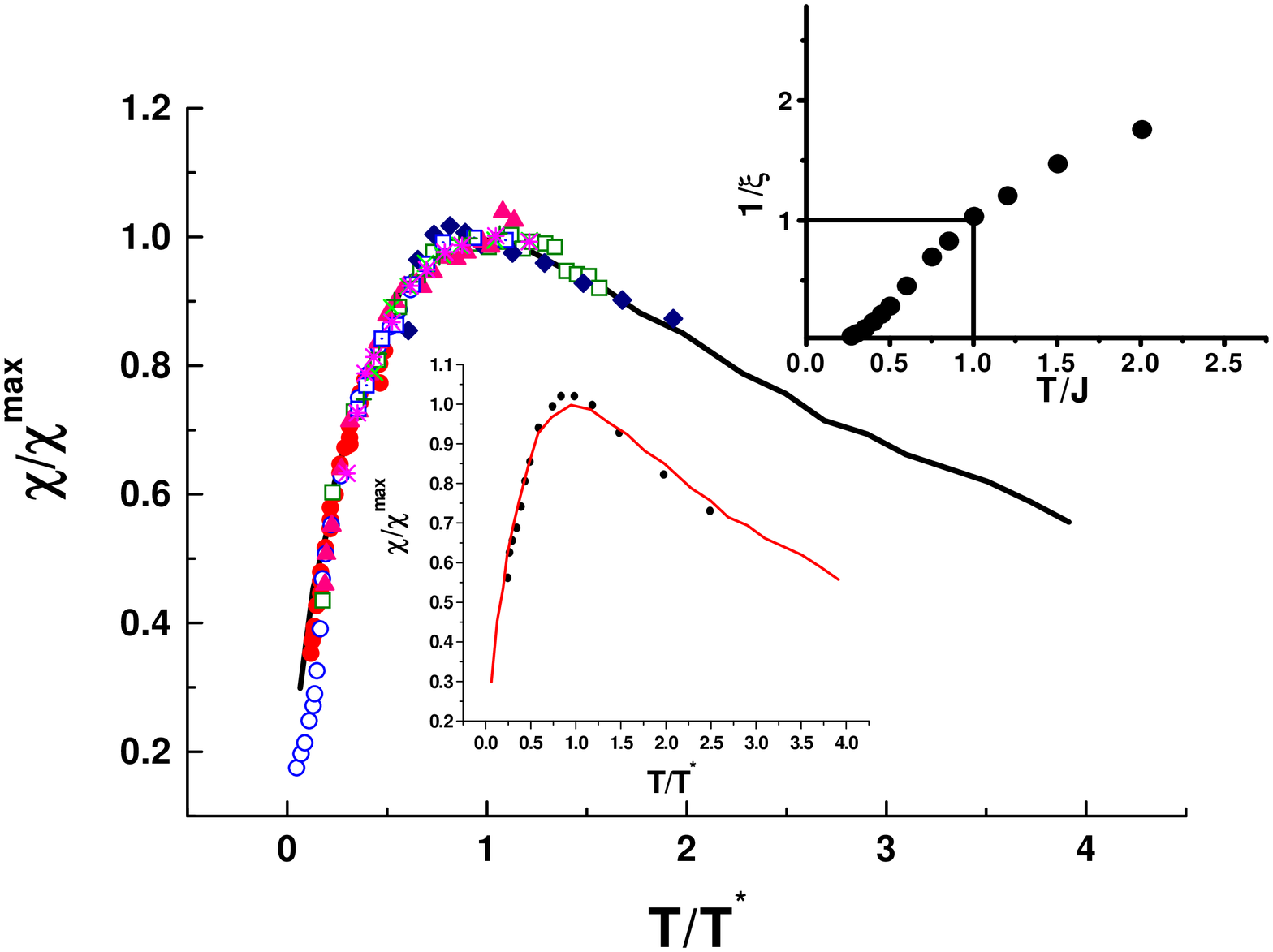}
\caption{Scaling for the $^{63}Cu$ Knight shift
 in YBa$_2$Cu$_3$O$_7$\cite{walstedt}, 
YBa$_2$Cu$_4$O$_8$\cite{curro97,zimmerman,bankay}, 
YBa$_2$Cu$_3$O$_{6.63}$\cite{takigawa}, and
Y$_{1-x}$Pr$_x$Ba$_2$Cu$_3$O$_7$ ($x=0.05,0.1,015,0.2$)\cite{reyes}, 
compared to the scaling function obtained
by Nakano \textit{et al.}\cite{nakano} (solid line) for the bulk spin susceptibility in La$_{2-x}$Sr$_x$CuO$_4$. 
$T^*$ is the temperature at which the Knight shift has a maximum. 
 The lower inset shows comparison of Nakano \textit{et al.}\cite{nakano} 
scaling function (solid line) to the Heisenberg model numerical calculations of Macivi\'c and Ding\cite{MD} (solid circles). 
The upper inset shows the numerical 
results for the  correlation  length\cite{MD}, which show that $\xi \simeq 1$ at temperature $T \simeq J$}
\label{knan}
\end{center}
\end{figure}

Almost 20 years after the discovery of high-$T_c$ cuprates,
there is still no consensus on the origin of superconductivity in these materials.  
In a large part this is due to an incomplete understanding of the normal state of matter
from which it arises. In the present communication we address this issue through a 
careful analysis of their measured low frequency magnetic properties.
We show that the scaling behavior found early on in the temperature-dependent component of the 
static uniform planar susceptibility for the 2-1-4 materials\cite{johnston,nakano} extends to the 1-2-3 materials,
and reflects the presence of a spin liquid that dominates the low frequency magnetic 
properties of the normal state from zero doping to doping levels of order $0.22$, the pseudogap region\cite{NPK}.
Quite remarkably, this universal part of the static response agrees very well with the theoretical
Monte Carlo calculations for the 2D Heisenberg model\cite{MD} (see Fig. \ref{knan}). There is, moreover, a second component
present in the Cu-O planes throughout this doping range; it is a Fermi liquid, whose temperature independent 
contribution to the uniform susceptibility is doping dependent\cite{johnston,nakano} because of the
doping-dependent changes in the relative fractions of spin liquid and Fermi liquid\cite{johnston,nakano,HS}. We are thus led to a two-component
phenomenological description of the low frequency magnetic properties in the pseudogap phase that provides a unified account of
their behavior, including the crossover seen in their dynamic properties\cite{imai,BP} at a temperature comparable to that 
at which the spin liquid susceptibility displays a maximum.
\begin{figure}
\begin{center}
\includegraphics[width=3in]{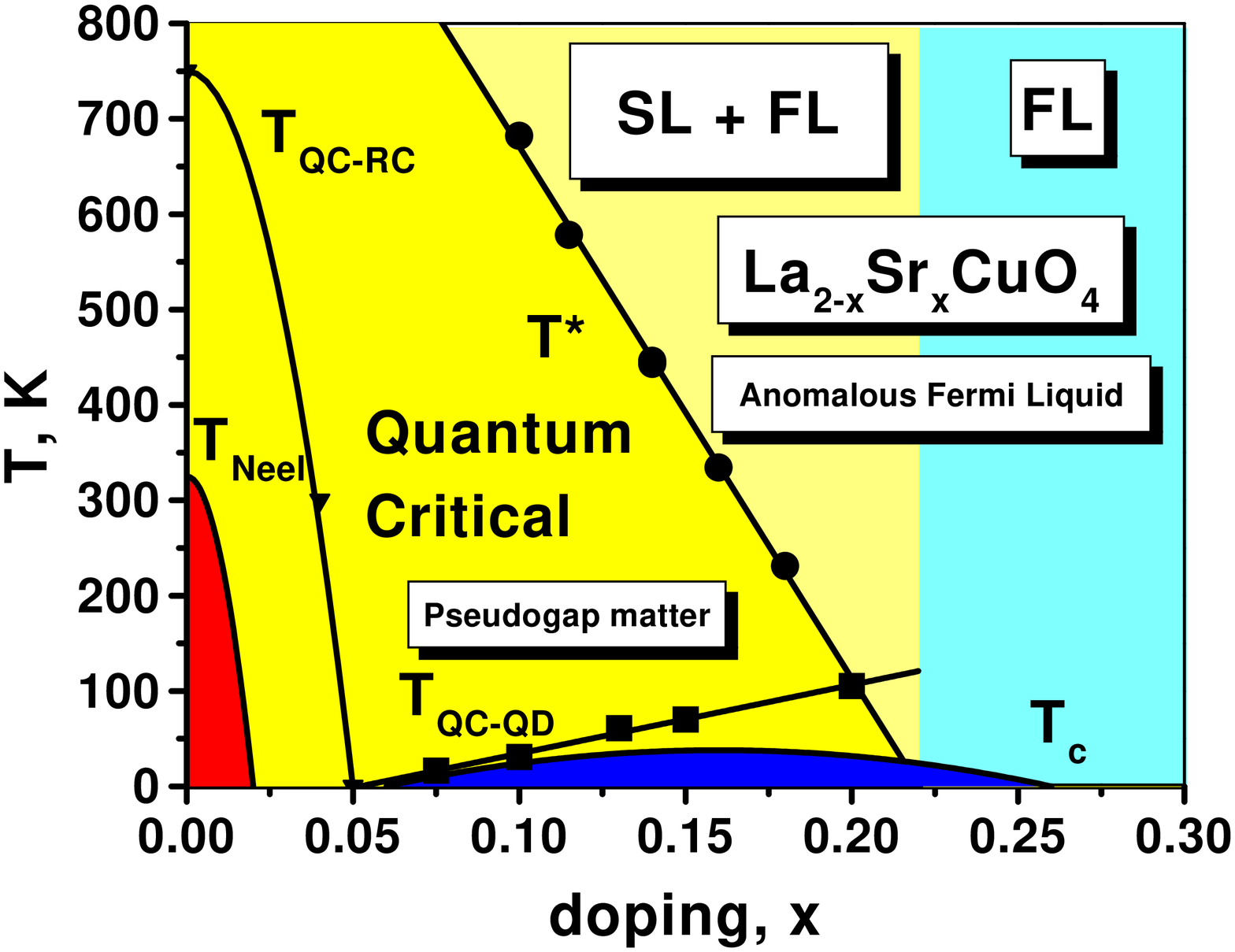}
\caption{The phase diagram for high-T$_c$ cuprate superconductors. $T^*(x)$ corresponds to a 
crossover  at $\xi(T,x)=1$ to the strongly correlated  phase in the spin liquid. $T_{QC-QD}$ is the spin liquid
crossover from the quantum critical (QC) to quantum disordered (QD), or gapped spin liquid, regime.
 $T_{QC-RC}$ is the spin liquid
crossover from the QC to renormalized classical (RC) regime.}
\label{phase}
\end{center}
\end{figure}
Our analysis suggests that in the superconducting state it is the doping dependent Fermi liquid component that 
forms the superconducting condensate, so that the doping dependence seen in the magnetic measurements will be 
directly reflected in the doping dependence of the superconducting condensate seen in penetration
depth and other experiments. We further find (see Fig. \ref{phase}), from our analysis of the spin-lattice relaxation
rate measured in NMR experiments on both the 1-2-3 and 2-1-4 materials that in the absence of superconductivity
there are two candidate quantum critical points (QCP) associated with the pseudogap matter. 
One suggested QCP is at a doping level $\sim 0.22$; 
it marks a $T=0$ quantum phase transition from the conducting Fermi liquid on the right-hand side (rhs) to pseudogap matter, 
in which a portion of the quasiparticle Fermi surface has lost its low frequency spectral weight, becoming localized in an insulating
 spin liquid that coexists with the remaining Fermi liquid component.
The second proposed QCP, previously identified by Ohsugi \textit{et al.} \cite{ohsugi}, is at $x \sim 0.05$ and marks a transition between two different
quantum states of the spin liquid; a quantum disordered phase on the rhs, which could be spin glass\cite{DP},and a magnetically ordered 
spin liquid on the left-hand side (lhs). The Fermi liquid is a 
bystander in this transition. 
There are several different ways in which such a transition could occur.
For example, the transition could be weak first order.
Therefore, we do not attempt the precise analysis of this question 
in the paper. Rather, we show the presence of two components 
in the spin dynamics at moderate doping levels, and note that
there is a change in its dynamics in the vicinity of a doping level of
$0.05$, which makes it a candidate for a QCP. The two-component spin
dynamics is we believe already present at high energies of the order of $J$.
 
We note that the presence of two critical points, and not just one, does not contradict 
finding a first order Mott transition, but rather confirms it. A proper thermodynamic variable for such
a transition would not be doping, but rather the chemical potential or pressure.
In some sense, doping is similar to volume for the first order liquid-gas transition, so
on general grounds one would expect a region of phase separation as a function of doping $x$, rather
than a sharp first-order transition. This point has been made by Lev Gor'kov in many publications\cite{GS,GT}.
Essentially our phase diagram implies the onset of a phase separation regime
at x=0.05 that persists up to  x=0.22, a viewpoint that is
confirmed by La NMR in La$_2$CuO$_{4+\delta}$\cite{hammel}. The lower critical
point corresponds to the onset of the phase separation behavior. 
At the second critical point the phase separation and the SL disappears.
Quite generally, it follows from the two-component picture that the transition to a gapped spin liquid can be 
either percolation type 
or stripe type, where the correlation length is limited by finite size of
the domain. The details and the energetics of how the
domain order appears at low dopings will determine the nature of the critical point at low doping level. 

We begin our analysis by writing the low frequency dynamic spin susceptibility in a simple two-component
form:
\begin{equation}
\chi(\bm{q}, \omega) = f(x) \chi_{SL}(\bm{q}, \omega) + [1-f(x)] \chi_{FL}(\bm{q}, \omega),
\label{dyn}
\end{equation}
where $f(x)$ is the fraction of spin liquid (SL), $1-f(x)$ that of a 2D Fermi liquid (FL).
In the static long wave length limit, the Fermi liquid bulk susceptibility, $\chi_{FL}$,
is both temperature and doping independent. When its contribution is subtracted from the
experimental data for the bulk spin susceptibility we find, as shown in Fig. \ref{knan},
that the 1-2-3 materials show the same scaling behavior with $T^*$, the temperature at which the spin
liquid bulk susceptibility is maximum, as had earlier been proposed for the 2-1-4 materials\cite{johnston,nakano}.
Not only is this scaling behavior universal, but $\chi_{SL}$ follows very 
well the calculated\cite{MD} bulk 
spin susceptibility for the Heisenberg model with a doping-dependent exchange constant 
$J(x) \simeq T^*(x)$. 
The spin liquid contribution,
\begin{equation}
f(x) \chi_{SL}(T) = \chi^{max} \tilde{\chi}(T/T^*(x)),
\end{equation}
is a universal function of $T/T^*$.  

A simple explanation for the scaling law found in the static susceptibility of the spin liquid is that as 
holes are added to the Cu-O plane, the hybridization of their orbitals with those of the localized Cu
spins makes it possible for the latter to hop. The resultant hopping reduces the global effectiveness of
the nearest neighbor interaction J. We note that the doping dependence of $T^*$ is not far from that predicted
by simple models\cite{PWA} of the consequences of that hopping,
\begin{equation}
T^* = J - tx, \ \ \ t \simeq 4.8J
\end{equation}
The spin liquid is dominant in determining the low frequency dynamic magnetic behavior and this dominance 
explains the success of the approach developed by Millis \textit{et al.}\cite{MMP} (hereafter MMP) in their analysis of the results
of NMR experiments on the cuprates, since the MMP dynamic susceptibility,
\begin{equation}
\chi_{SL}(\bm{q},\omega) = \frac{\chi_{\bm{Q}}}{1 + \xi^2 (\bm{q} - \bm{Q})^2 - i \frac{\omega}{\omega_{SF}}\,}\,,
\label{SLpar}
\end{equation}
is that appropriate for the scaled 2D Heisenberg nearly antiferromagnetic liquid of localized spins characterized
by their peak static susceptibility, 
\begin{equation}
\chi_{\bm{Q}} = \alpha \xi^2,
\end{equation} 
a correlation length, $\xi$, and a relaxational frequency, $\omega_{SF}$. $\alpha$ is a temperature-independent constant.

NMR measurements of the $^{63}Cu$ spin-lattice relaxation rate, $^{63}T_1$,
and the spin-echo-decay time, $^{63}T_{2G}$, have shown that below $T^*$ one is in the $z=1$
dynamic scaling regime\cite{BP,imai} expected for the quantum critical (QC) regime of a 2D
antiferromagnet, so that, as discussed in Ref.\cite{BP}, one can describe the spin dynamics using the quantum 
non-linear sigma model, or spin wave theory\cite{CHN,CSY}. The resulting QC scaling theory\cite{CSY} for the spin
liquid without long-range order gives a linear dependence of the correlation length on temperature,
\begin{equation}
\frac{1}{\xi(T,x)}\, = 1.04 \frac{T}{c}\, + a(x),
\end{equation}
where $c \propto J \sim T^*$ is the spin wave velocity, and the offset $a(x) > 0$ goes to zero at $x_c$, a QCP that marks the onset of long-range order.
It yields a similar linear dependence on $T$ for 
$^{63}T_1T \propto \omega_{SF}$\cite{BP,imai},
\begin{equation} 
^{63}T_1T = A(x) + \kappa T^* \frac{T}{T^*},
\end{equation}
with an offset $A(x)$ that measures the distance from the proposed QCP. 
\begin{figure}
\begin{center}
\includegraphics[width=3in]{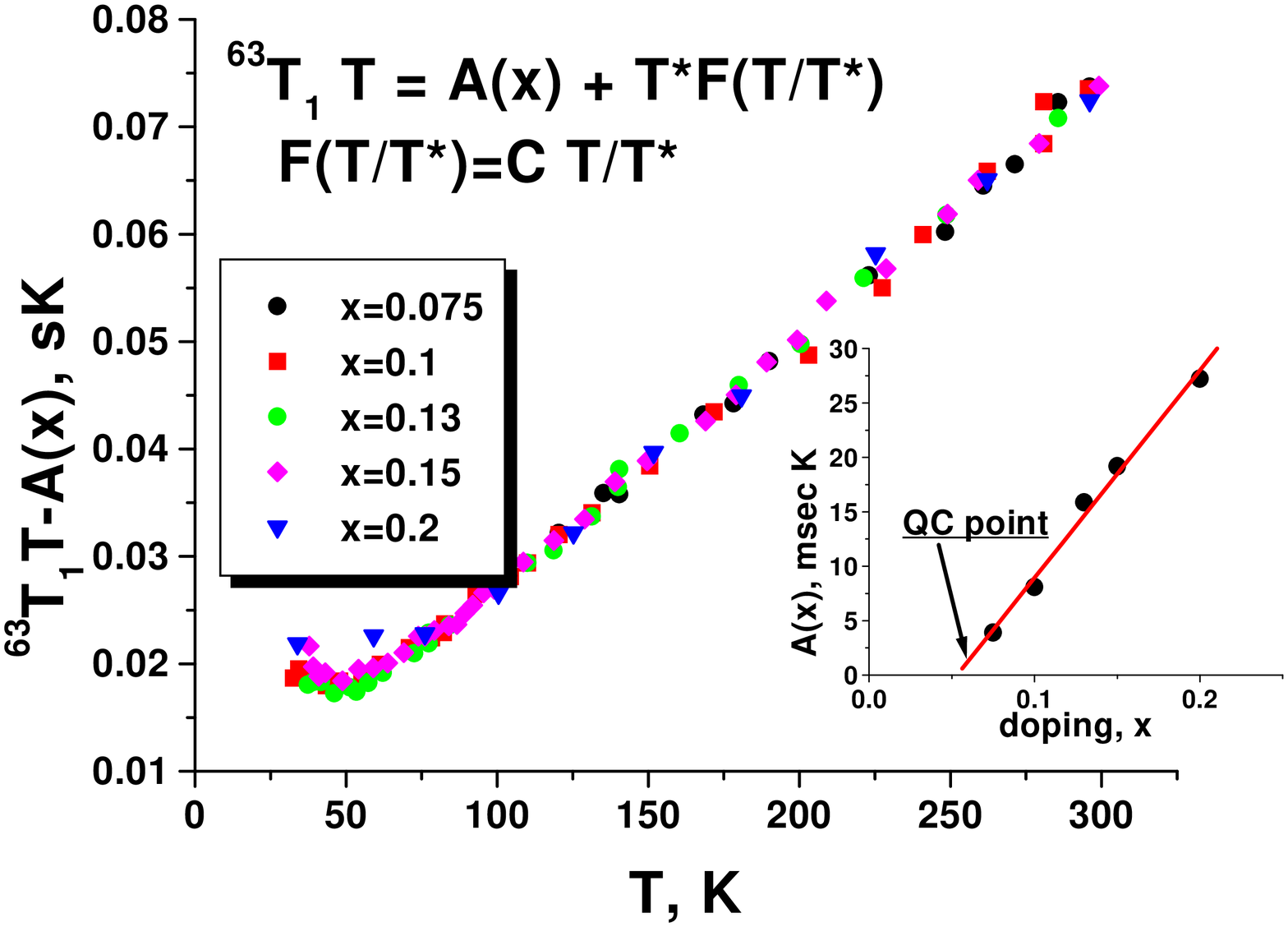}
\includegraphics[width=3in]{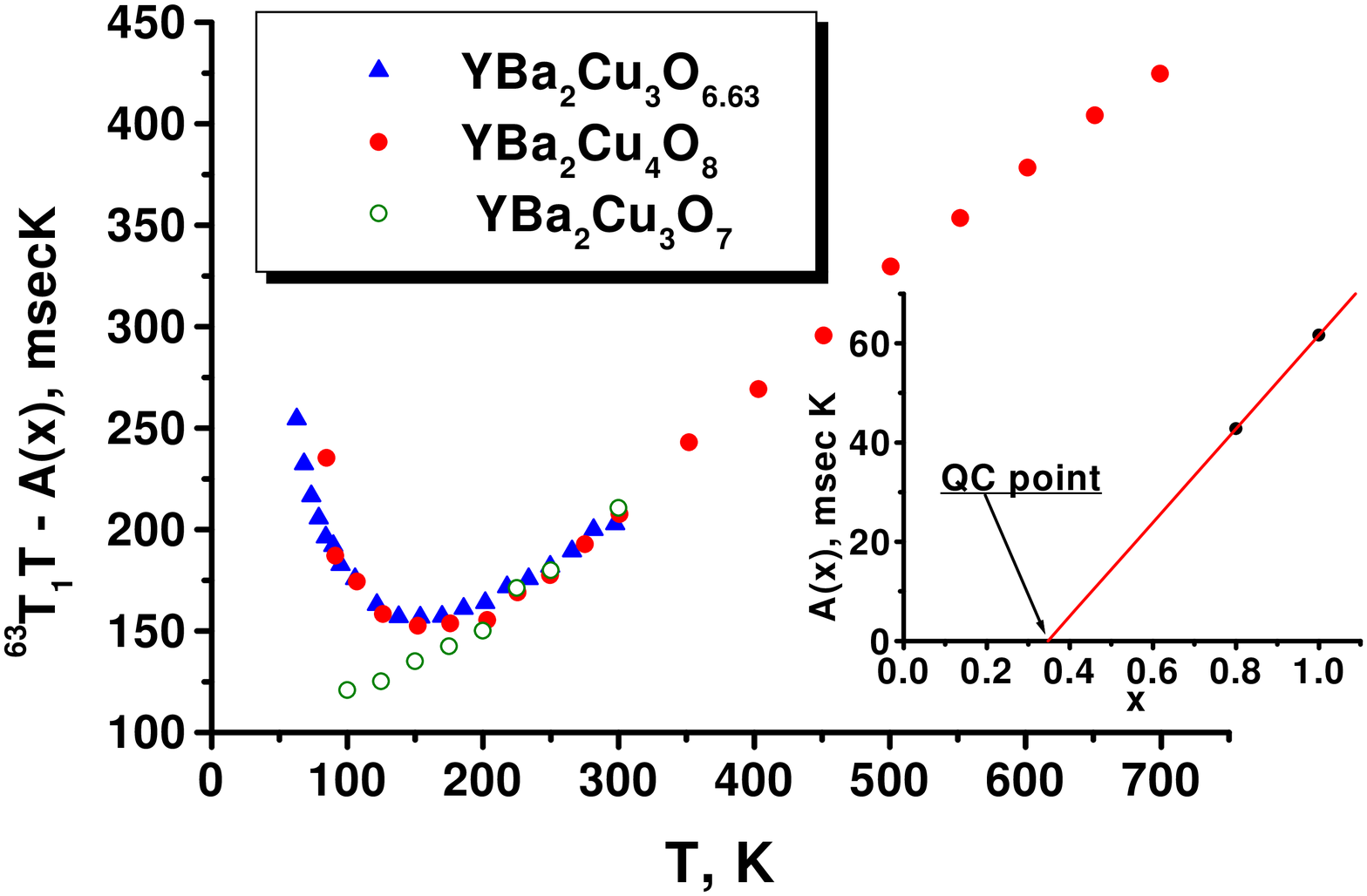}
\caption{Scaling for the NMR relaxation rate $T_1$ (a) in La$_{2-x}$Sr$_x$CuO$_4$\cite{ohsugi} (b) in YBa$_2$Cu$_3$O$_{6+x}$. 
The offset $A(x)$ shown in the insets depends linearly on $x$, and points to a
possible QCP in the spin liquid.}
\label{T1T}
\end{center}
\end{figure}
The data collapse for both
families shown in Fig.\ref{T1T} suggests this QCP is at $x \sim 0.05$.
A similar analysis can be done for the high-temperature $^{63}T_1$ 
data\cite{imai} on the insulating side; it results in the same critical point, as shown on the phase diagram in
Fig. \ref{phase}.
 
The QCP at $x \sim 0.05$ corresponds to a zero-temperature phase transition to a
magnetically ordered state at $x<0.05$, which could be a 2D antiferromagnet, or a spin glass.
A finite correlation length at $T \neq 0$ or $x>0.05$ implies the presence of a gap in the 
spin excitation spectrum,
\begin{equation}
\epsilon(\bm{q}) = \sqrt{\Delta^2 + c^2 (\bm{q}-\bm{Q})^2}, \ \ \Delta = \frac{c}{\xi(T,x)}\,.
\end{equation}
Here $c$ is the spin wave velocity. In the $z=1$ QC regime the gap $\Delta$
tracks $\omega_{SF}$. For $x > 0.05$ the gap $\Delta$ saturates at low temperatures to
$\Delta_0 = c/\xi_0(x)$ at the crossover to the quantum disordered (QD) gapped spin liquid regime. 
The constant energy gap $\Delta_0$  results in an exponential decrease of damping with 
temperature in the spin liquid, or, equivalently, an exponential increase of $\omega_{SF}$. The energy
gap can appear in the 2D quantum spin liquid\cite{CHN,CSY}, or in 1D stripes of spin liquid\cite{tranquada2} due to 
reduced dimensionality. Our analysis does not distinguish between these possibilities. 
The QC-QD crossover explains the sharp upturn in $^{63}T_1T$ at low temperatures 
observed in NMR experiments on 1-2-3 materials. Since $c(x) \propto T^*(x)$, and decreases linearly to zero at $x=0.22$, 
while $1/\xi_0(x)$ increases linearly from $x=0.05$, the zero-temperature energy gap $\Delta_0(x)$ will vanish
at \textit{both} critical points, and will have a bell-shaped form similar to $T_c(x)$.  

Our analysis also casts light on the issue of intrinsic spatial inhomogeneity in the pseudogap state\cite{tranquada1,tranquada2,emkivf}. 
The issue is how to reconcile the commensurate spin susceptibility peaks required by the spin liquid and the fit to the NMR 
data\cite{walst},
with the incommensurate peaks seen in inelastic neutron scattering experiments\cite{aeppli}. 
The resolution was suggested some time ago by Slichter\cite{slichter}, who pointed out that if, 
as a result of Coulomb interaction between the holes, regions that were hole rich formed domain walls between regions
in which one had nearly antiferromagnetic commensurate behavior, the two kinds of experiments were compatible.
This issue has been independently confirmed and clarified by 
Tranquada \textit{et al.}\cite{tranquada1,tranquada2}, who 
found that the spin wave dispersion seen in cuprate superconductors by inelastic neutron scattering 
is consistent with commensurate spin response of a spin ladder, while the incommensurate structure appears
as a result of the fluctuating stripe order. 
Further significant experimental evidence of such order is provided by the experimental
observation of two incommensurate peaks instead of four in a detwinned sample of YBa$_2$Cu$_3$O$_{6.6}$\cite{mook}.

The above results make clear the key role played by the spin liquid component in the pseudogap phase. 
What of the Fermi liquid component? Our analysis strongly suggests that its 
superconductivity must arise from its coupling to the magnons that are the excitations of 
the spin liquid. 
The superconducting mechanism is thus electron-magnon
coupling which gives rise to an effective  magnetic quasiparticle
interaction responsible for the measured d-wave pairing state and
superconductivity at high temperatures.
The presence of intrinsically electronic dynamic inhomogeneous
behavior is, to some extent, incompatible with the existence of a sharp
Fermi surface, so, in the pseudogap phase it is not clear how one 
develops a consistent mathematical description of the formation of the superconducting condensate.
Some guidance in this regard is provided by considering doping levels that exceed 0.22, since in this magnetically
overdoped region the spin liquid component disappears, leaving behind a uniform Fermi liquid with strong magnetic
quasiparticle interactions capable of bringing about a high $T_c$\cite{MP}. Indeed, starting from the magnetically overdoped
side, and reducing the doping level, it is appealing to assume that one still has a mostly intact Fermi surface; only
those quasiparticles in the vicinity of the hot spots coupled most strongly by magnetic interactions form the spin liquid, and 
in the process lose their low frequency spectral weight. In this scenario, the remaining ``cold'' quasiparticles are
those capable of forming the condensate, while in the normal state these can be seen in ARPES experiments\cite{norman} as Fermi
arcs. Again, this picture likely breaks down once an appreciable part of the FS has been transformed to a spin liquid.

In conclusion, the underdoped pseudogap phase in cuprate superconductors is a new state of matter, which arises at hole
doping level below $x=0.22$. We have shown that this phase is a mixture of the Fermi liquid phase, which
occurs at high doping levels, and the spin liquid phase of the parent insulating compound.
Our results are presented on the phase diagram for the pseudogap state shown in Fig. \ref{phase}. 
Two quantum critical points are evident from our analysis of available experimental data. 
One critical point, located at the hole doping $x \simeq 0.05$, is spin liquid only, and
corresponds to the disappearance of the long-range magnetic order. 
The second, hidden critical point is located at the hole doping level of $x \simeq 0.22$, where 
the spin liquid fraction vanishes. It turns out that the spin wave velocity and $T^*(x)$ for the spin liquid component
also vanish at  the second critical point. 

We would like to thank  N. Curro, L.P. Gor'kov,
J. Haase, J. Schmalian, and C.P. Slichter
for stimulating discussions, and acknowledge support from the US Department of Energy.

\end{document}